\documentstyle[preprint,eqsecnum,aps]{revtex}
\newcount\binary   \binary=0
\newcount\precount   \precount=0
\newcount\zero   \zero=0
\font\tenmext=cmex10 scaled\magstep2
\mathchardef\rbigbrace="7329
\mathchardef\rmedbrace="736F
\mathchardef\rsmabrace="7309
\mathchardef\rtop="7339
\mathchardef\rmid="733D
\mathchardef\rlin="733E
\mathchardef\rbot="733B
\def\putlargebrace{${{{{{\rtop\atop\rlin}
\atop \rmid }\atop
\rlin} \atop\rbot}}$
\hskip-4.28ex\raise5.7ex\hbox{$\rlin$}
\hskip-3.10ex\raise4.5ex\hbox{$\rlin$}
\hskip-3.09ex\raise-0.2ex\hbox{$\rlin$}
\hskip-3.09ex\raise-1.4ex\hbox{$\rlin$}
}
\def\puthugebrace{${{{{{{{{{\rtop\atop\rlin}
\atop \rlin}\atop \rlin} \atop \rmid }\atop
\rlin}\atop \rlin}\atop\rlin} \atop\rbot}}$
\hskip-4.70ex\raise9.0ex\hbox{$\rlin$}
\hskip-2.69ex\raise8.1ex\hbox{$\rlin$}
\hskip-2.675ex\raise7.2ex\hbox{$\rlin$}
\hskip-2.635ex\raise6.3ex\hbox{$\rlin$}
\hskip-2.685ex\raise2.0ex\hbox{$\rlin$}
\hskip-2.650ex\raise1.1ex\hbox{$\rlin$}
\hskip-2.645ex\raise0.2ex\hbox{$\rlin$}
\hskip-2.685ex\raise-0.7ex\hbox{$\rlin$}
}
\def\putmammothbrace{${{{{{{{{{{{{{{{{{{{\rtop\atop\rlin}
\atop \rlin}\atop \rlin}\atop  \rlin}\atop  \rlin}\atop  \rlin}\atop
 \rlin}\atop \rlin}\atop \rmid }\atop
\rlin}\atop  \rlin}\atop \rlin}\atop \rlin}\atop \rlin}\atop
\rlin}\atop \rlin}\atop\rlin} \atop\rbot}}$
\hskip-7.22ex\raise18.2ex\hbox{$\rlin$}
\hskip-2.68ex\raise17.2ex\hbox{$\rlin$}
\hskip-2.63ex\raise16.2ex\hbox{$\rlin$}
\hskip-2.67ex\raise15.6ex\hbox{$\rlin$}
\hskip-2.615ex\raise14.7ex\hbox{$\rlin$}
\hskip-2.705ex\raise13.8ex\hbox{$\rlin$}
\hskip-2.68ex\raise12.9ex\hbox{$\rlin$}
\hskip-2.69ex\raise12.0ex\hbox{$\rlin$}
\hskip-2.625ex\raise11.1ex\hbox{$\rlin$}
\hskip-2.665ex\raise6.9ex\hbox{$\rlin$}
\hskip-2.63ex\raise6.0ex\hbox{$\rlin$}
\hskip-2.71ex\raise5.2ex\hbox{$\rlin$}
\hskip-2.70ex\raise4.3ex\hbox{$\rlin$}
\hskip-2.59ex\raise3.4ex\hbox{$\rlin$}
\hskip-2.71ex\raise2.5ex\hbox{$\rlin$}
\hskip-2.61ex\raise1.6ex\hbox{$\rlin$}
\hskip-2.69ex\raise0.7ex\hbox{$\rlin$}
\hskip-2.69ex\raise-0.2ex\hbox{$\rlin$}
}
\def\putbigbrace{\raise6.25ex\hbox{$\rbigbrace$}}
\def\putmedbrace{\raise2.75ex\hbox{$\rmedbrace$}}
\def\putsmabrace{\raise1.75ex\hbox{$\rsmabrace$}}
\begin{document}
\draft
{\ifpreprintsty \global\advance\precount by 0
          \else \global\advance\precount by 1 \fi}
\tightenlines \global\advance\binary by 1
\preprint{
\rightline{\vbox{\hbox{\rightline{MSUCL-937}}
\hbox{\rightline{nucl-th/9410028}}}}
         }
\title{Collision rates for $\bbox{\rho}$, $\bbox{\omega}$ and
$\bbox{\phi}$ mesons at nonzero temperature}
\author{Kevin Haglin\cite{myemail}}
\address{
National Superconducting Cyclotron Laboratory, Michigan
State University\\
East Lansing, Michigan 48824--1321, USA
}
\date{\today}
\maketitle
\begin{abstract}
The $\omega$ and $\phi$ mesons are exceptionally narrow in vacua being
about 8 and 4 MeV.   At finite temperature they will scatter with
other hadrons in their approach to equilibrium and thereafter.  These
dynamics induce broadening to the already unstable vector mesons which
is calculated in the framework of relativistic kinetic theory.  A rather
complete set of low-lying mesons provide reaction partners whose interactions
are modeled using effective Lagrangians.  Collision rates for $\rho$ and
$\omega$ are found to be $\sim$ 100 MeV at $T$=200 MeV while the rate for
$\phi$ is $\sim$ 25 MeV.  Corresponding mean free paths are about 1 and 4 fm!
Possible observable consequences are discussed.
\end{abstract}

\narrowtext

\section{Introduction}
\label{sec:intro}

Vector meson behaviour at nonzero temperature is particularly important for
systems created in ultra-relativistic heavy-ion collisions.  Details of
their creation, propagation and decay greatly affect the overall many-body
dynamics of the collisions and consequently the observed particle spectra.
Pion or kaon correlations, strangeness production and absolute normalization
of dilepton spectra are just a few features deeply linked
to the way in which vector mesons behave in hot, strongly interacting matter.
Primary emphases in past studies on this topic have been on possible mass
shifts, modified dispersion relations and modifications to the
widths\cite{cdet85,sgot87,mdey90,rfur90,cgal91,cson93,klhcg}.  Results for
isoscalars and certain neutral isovectors have direct experimental
significance since at minimum they might be observed through their decays
into lepton pairs.

The $\rho^{0}$ meson decays strongly into oppositely charged pion pairs with a
lifetime in the vacuum of $\sim$ 1.3 fm/$c$.  The $\omega$ decays less often
with a vacuum lifetime $\sim$ 23 fm/$c$ and mostly into the pion triplet
$\pi^{+}\pi^{-}\pi^{0}$ owing to its negative G-parity.  Some difficulties
would surely arise distinguishing one from the other in dilepton channels
since they are nearly degenerate in mass---unless some spectacular differences
in medium arose.  The $\phi$, on the other hand, is well removed with a
vacuum mass above 1 GeV.  It decays primarily into kaon pairs and $\pi\rho$
combinations with average lifetime $\sim$ 44 fm/$c$ in the vacuum.  At
temperatures of the pion mass or above, broadening to a ``full'' width
occurs in each of these mesons due to scattering since densities are nearly
1 hadron/fm$^{3}$.  Even rather nonreactive hadrons like the $\phi$ meson will
scatter.  Are the mean free paths for these particles about 1 fm or perhaps
something much greater?  Should any of them be expected to pass completely
through reaction zones without interacting at all?  The transverse size of
the zone varies as $R =$ 1.2 fm $\times$ $A_{p}^{1/3}$ where $A_{p}$ is the
projectile mass number.  For the heaviest beams expected in future experiments
$R$ is about 10 fm.  Different outcomes could be expected depending on
whether the average collision distances for the vector mesons are larger,
nearly the same, or smaller than $R$.

In heavy-ion collisions at high energies the nuclei pass completely
through each other and there is for a short time left behind a highly
excited, very hot system of matter.  Here is where one might expect
to have quarks and gluons (partons) relatively free from nonperturbative
effects.  The partons are able to traverse distances larger than typical
hadron sizes and one speaks of the quark-gluon plasma (QGP).  Tremendous
progress has been made recently in understanding some of the possible
details of the QGP.  A few specifics include thermalization time
scales\cite{klaus1}, chemical equilibration\cite{kgjk}, particle
production\cite{klaus2}, temperature scenarios\cite{shuryak}, and the role
of minijets\cite{wang}.  The system will eventually crossover from the plasma
phase into a purely interacting hadronic state.  Popular guesses
as to where this occurs in temperature are near $200$ MeV.  Then for
several fm/$c$ the system will interact strongly, cooling and expanding as
it goes.  Time and distance scales of interaction for the lightest hadrons
are 1 to perhaps a few fm\cite{khsp}.  It cools and the pressure
drops, expansion takes over and collisions cease to occur sometime about
30 or more fm/$c$ later, depending on such details as the roles of expansion
and nucleation and subsequent degree of supercooling\cite{joe&laszlo}.  At
this point the constituent hadrons (mostly pions) free
stream.  Understanding of the high temperature domain for the hadron phase
has also improved through recent studies.  Finite temperature behaviour of
pions, $\rho$, $\omega$, $\phi$ and $a_{1}$ mesons has been
considered\cite{cgal91,klhcg,song}.  Temperature's effect has been
investigated using finite-temperature field theory at the one-loop level for
these mesons.  But one of the basic sources of possible modification from
vacuum properties occurs at the two-loop level and has been largely
ignored---that of pure collisional modification.  With regard to the $\phi$
meson Ko and Seibert\cite{seko} recently reported finding the full width
including scattering was less than 10 MeV for all temperatures.  But they
neglected some important contributions which are included here and
furthermore, their results were somewhat sensitive to the ``in-medium'' width
of propagating kaons which was fixed first at 20 MeV and then 30 MeV for
comparison.   Improvements to this aspect of the calculation are made here
by introducing a temperature and energy dependence to the widths of the
exchanged particles.

This paper contains results of collisional effects on $\rho$, $\omega$ and
$\phi$ mesons and is organized as follows.  In the next section a model for
the interacting hadronic system is discussed and its constituent degrees of
freedom enumerated.  Interactions are modeled with standard, yet simple
Lagrangians consistent with a more elaborate and complete effective chiral
Lagrangian.  Coupling constants are chosen to give consistency with observed
decays.  Strong-interaction form factors are presented and briefly discussed.
For some kinematic configurations in scattering the exchanged particles in
$t$-channel graphs can become timelike and even approach mass shell.  Pions
and kaons are affected in this study and consequently acquire imaginary parts
in their propagators which are estimated by computing temperature and momentum
dependent collision rates.  For simplicity, Breit-Wigner cross sections are
used to account for interaction with pions, kaons and $\rho$ mesons in these
estimates.  Section~\ref{sec:rates} contains the kinetic theoretic results for
time and distance scales of interaction.  Collision rates for all three
vector mesons are computed first and then approximate mean free paths are
obtained.  Results are shown as a function of temperature in the range $100
\le T \le 200$ MeV.  A short discussion and summary follows in
Sec.~\ref{sec:summary} highlighting possible consequences for $\rho$,
$\omega$ and $\phi$-related observables.

\section{Modeling the Hadronic Fields}
\label{sec:model}

One might model the hadronic cooling stage as a thermal ensemble of
low-lying hadrons: pions, kaons, $\rho$ and $K^*$(892) mesons as
well as some heavier axial-vectors $a_{1}(1260)$, $b_{1}(1235)$ and
$K_{1}(1270)$.   The $\phi$ meson resides here also and like the others
will be assumed to be thermalized.  A posteriori this should be checked against
average collision distances for validity.  These particles are quantified as
fields which interact according to the following Lagrangians\cite{um88,gj94}
\begin{eqnarray}
L_{VPP} &=& g_{VPP}\, V^{\mu} P \stackrel{\leftrightarrow}
{\partial_{\mu}} P \nonumber\\
L_{VVP} &=& g_{VVP}\,
\epsilon_{\mu\nu\alpha\beta}\, \partial^{\mu}V^{\nu}
\partial^{\alpha}V^{\beta}\,P \nonumber\\
L_{AVP} &=& g_{AVP}\, A_{\mu\nu}V^{\mu\nu} \,P, \nonumber\\
\label{eq:lagrangians}
\end{eqnarray}
where the field strength tensors are $A_{\mu\nu} = \partial_{\mu}A_{\nu}-
\partial_{\nu}A_{\mu}$ and similarly for $V^{\mu\nu}$.  In the above
expressions $A, P$ and $V$ refer to axial-vector, pseudoscalar and vector
meson fields.  Elementary processes can now be considered.  To be completely
general at this stage, consider the two diagrams shown in
Fig.~\ref{fig:diagram}, where the species $V$ will be a $\rho$, $\omega$
or a $\phi$ and the set $\{a, 1, 2, R, E\}$ are as-of-yet unspecified.
Since hadronic cross sections tend to be dominated by resonances, one
naturally looks to $\rho\pi$ scattering through $a_{1}(1260)$, $\rho K$
scattering through $K_{1}(1270)$, $\omega \pi$ scattering through
$b_{1}(1235)$ and to strange-particle possibilities for the $\phi$.  Although
it can form $K_{1}(1770)$ and $K_{4}(2045)$ when
scattering with kaons and $K^{*}$\,s, the branching ratios are very small
leaving them inconsequential.  There are also resonance possibilities in the
non-strange sector.  Scattering with pions is probable while forming
$b_{1}(1235)$ or even $\rho(1450)$ mesons.  The branching ratios are not firmly
established, but tentative upper limits of 1.5 and 1 percent have been
quoted\cite{pdg}.  Then in the $t$ channel there are several seemingly equally
important possible configurations since all three vector mesons can
scatter with pions, kaons and a host of others.  In considering diagrams for
the general reaction with any of the vector mesons two problematic components
immediately arise, namely, propagators and form factors.

Phase space allows the squared four-momentum of the exchanged particle
to access timelike values and even become equal to its mass squared for
those processes in which $m_{1}+m_{E} < m_{V}$.  If $t$ is identically equal
to $m_{E}^{2}$ then one has a true on-shell decay instead of scattering.
This must be interpreted carefully so as to avoid double counting. The
finite-temperature propagator at tree-level for the exchanged particles
should, in principle, have the regular term $1/(p^{2}-m^{2})$ plus a term
which goes like $2\pi \delta(p^{2}-m^{2})n_{\beta}$\cite{jfdbrh}.  If the
second term is ignored while at the same time an imaginary part is introduced
into the denominator of the first one by hand in order to account for the
effect of the matter, i.e. $1/(p^{2}-m^{2}-im\Gamma)$, the result consistently
accounts for scattering although somewhat phenomenologically.  This is not
meant to be a substitute for an evaluation of the two-loop diagrams which
would consistently account for scattering plus higher-order corrections to
decays at finite temperature, but instead an estimate of it.   The appeal of
the kinetic theory approach is that it is very straightforward and yet
roughly captures the effect of the hot matter.

Hadrons necessitate the use of form factors since they are
composite objects and have finite extent which can be seen when probed with
higher and higher momentum transfers.  A suppression of this region is
physically imperative.  The standard way to accomplish this is to insert
a monopole form factor
\begin{equation}
F_{\alpha}(t) = {\Lambda^{2}-m_{\alpha}^{2} \over
\Lambda^{2} -t }
\label{eq:ffactor}
\end{equation}
at each vertex in a $t$-channel diagram where ${\alpha}$ indicates
a species for the exchanged particle.  Besides the $\rho$-$\pi$-$\pi$
vertices in the reaction
$\rho+\pi\to \rho+\pi$ which requires special attention to be discussed in a
moment, Eq.~(\ref{eq:ffactor}) is adopted
and a value of $\Lambda$ = 1.8 GeV is taken for all vertices.  Some
uncertainty is duly noted since this parameter should
really be species-dependent but in the absence of data with which one might
compare this prescription is at least a reasonable approximation of the common
finite-size features.  The monopole form is normalized so that when the
intermediate particle is on shell the form factor goes to 1.  For
most processes in meson-meson scattering this normalization is respected
and $F > 1 $ regions are kinematically inaccessible.  However, the intermediate
pion in the $t$ channel of the reaction $\rho+\pi\to\rho+\pi$ is able to
access this ``hyper-normalized'' region.  To generalize the monopole form
factor in a manner which is properly normalized when applied to the present
case it is taken to be
\begin{equation}
\tilde{F}_{\pi}(t) = {\Lambda^{2} - t_{\rm max} \over \Lambda^{2} - t}
\label{eq:fgactor}
\end{equation}
where the maximum momentum transfer is
\begin{equation}
t_{\rm max} = m_{\rho}^{2}+m_{\pi}^{2}-{1\over 2s}\left((s+m_{\rho}^{2}
-m_{\pi}^{2})(s+m_{\pi}^{2}-m_{\rho}^{2})-\lambda(s,m_{\rho}^{2},
m_{\pi}^{2}) \right).
\end{equation}
The kinematical ``triangle function'' is $\lambda(x,y,z)$ = $x^{2}-2x(y+z)
+(y-z)^{2}$\cite{byckling} and a value $\Lambda = m_{\rho}$ is taken to soften
the vertices.

Coupling constants naturally appear in the scattering amplitudes.  They are
adjusted in order to obtain the observed partial decay rates.  Relevant rates
for general decays are computed starting from the interactions in
Eq.~(\ref{eq:lagrangians}) to be
\begin{eqnarray}
\Gamma_{V\to PP^{\prime}} &=& {g_{VPP^{\prime}}^{2} \over 6\pi}
\left({|\bbox p\,|^{3} \over m_{V}^{2}}\right), \nonumber\\
\Gamma_{V\to PV^{\prime}} &=& {g_{VPV^{\prime}}^{2} \over 12\pi}
|\bbox p\,|^{3}, \nonumber\\
\Gamma_{A\to VP} &=& {g_{AVP}^{2} \over 6\pi}
|\bbox p\,| \left({\left(m_{A}^{2}+m_{V}^{2}-m_{P}^{2}\right)^{2}
\over 2m_{A}^{2}}+m_{V}^{2} \right),
\label{eq:elemdecay}
\end{eqnarray}
where $\bbox{p}$ is the center-of-mass momentum of the decay products.
{}From here one can get all the necessary coupling constants.  They are
listed in Table~\ref{table:one}.  Notice that as they are used in the
Lagrangians of Eq.~(\ref{eq:lagrangians}), some are dimensionless and
some carry units of inverse mass.

As mentioned already, timelike and intermediately propagating pions or kaons
acquire imaginary parts due to the presence of the matter which are
approximated by computing the dominant scattering rates.  For instance, pions
will scatter with other pions---a process dominated by the $\rho$ resonance.
They will also scatter with kaons to form $K^{*}(892)$ resonances.  The rate
at which each of these happens is roughly the density times the cross section
times the relative velocity.  More precisely, the contribution to the
collision rate of particle $b$ from pions is
\begin{equation}
\Gamma^{\rm coll}_{b} (E,T) =
\int\, ds \, {d^{3}p_{\pi}\over (2\pi)^{3}}
f_{\pi} \sigma_{b\pi}(s) v_{rel} \delta\left(s-(p_{b}
+p_{\pi})^{2}\right),
\label{eq:gammaofe}
\end{equation}
where $f$ is the Bose-Einstein distribution function and
\begin{equation}
v_{rel} = {\sqrt{(p_{b}\cdot p_{\pi})^{2}-4m_{b}^{2}m_{\pi}^{2}}
\over E_{b}E_{\pi}}.
\end{equation}
A Breit-Wigner form for the cross section
\begin{equation}
\sigma_{b\pi}(\sqrt{s}) = \left(2J_{\rm res}+1\right)
{\pi \over {\bbox{k}}^{2}}
{\Gamma_{{\rm res}\rightarrow b\pi}^{2} \over (\sqrt{s}-m_{\rm res})^{2}
+\Gamma_{\rm res}^{2}/4}
\end{equation}
is used with ${\bbox{k}}$ being the center-of-mass momentum, ``res''
refers to the resonance through which it proceeds and $b$ refers generally
to pions or kaons.   In the case of kaons, resonances $K^{*}$ and
$K_{1}(1270)$ are allowed since scattering with pions and with $\rho$ mesons
are frequent.  The width is a function of the three-momentum separately
and is therefore frame-dependent but determined by the hadron gas.
Results for pions and kaons are presented in Figs.~\ref{fig:piandkawidth}$a$
and $b$ respectively, each for temperatures 100, 150 and 200 MeV.  Depending
on the temperature and on the pion or kaon momenta, widths vary from 20 to
200 MeV.  Results similar to these were recently obtained for the momentum
dependent scattering rate of $\omega$ mesons at finite temperature using this
prescription\cite{kh94}.  These broad widths
could drastically modify the propagators and consequently modify the
vector-meson reactions.  Propagators for intermediate pions and kaons,
indicated generally by species $b$, will henceforth be
\begin{equation}
i \Delta_{b}(p^{2},E,T) = {i \over p^{2}-m_{b}^{2} - im_{b}\Gamma_{b}(E,T)}.
\label{eq:prop}
\end{equation}

\section{Time and distance scales of interaction}
\label{sec:rates}

The hot matter created in heavy-ion collisions of the type considered here
is populated mostly by pions and kaons and one knows their collision times
are $\sim 1$ fm/$c$ at high temperatures\cite{welke,weldon,goity}.  The system
is temporarily supported by its own collisions.
Thermodynamic estimates of pion and kaon abundances are 0.3 and 0.1 per
fm$^{3}$ for high temperatures while they are 0.15, $4.7\times 10^{-2}$ and
$ 1.9\times 10^{-2}$ per fm$^{3}$ for $\rho$, $\omega$ and $\phi$ mesons
at $T=200$ MeV.  It is interesting to see just how often they will collide
in this model for the reaction zone in the moments before freezeout.

\subsection{Collision Rates}

For a general (bosonic) reaction involving the vector $V$ and having the form
$V+a\rightarrow 1+2$, the kinetic theory expression for an
average scattering rate is
\begin{eqnarray}
{\overline{\Gamma}^{\, \rm coll}_{V}} &=&  {1\over n_{V}}
{d_{V}d_{a}} \int d^{3}\bar{p}_{V}d^{3}\bar{p}_{a}d^{3}
\bar{p}_{1}d^{3}\bar{p}_{2}
|\bar{\cal M}|^{2} \, (2\pi)^{4} \delta^{4}\left(p_{V}+
p_{a}-p_{1}-p_{2}\right)
\nonumber\\
& &\quad\quad\times f_{V}f_{a}\left(1+f_{1}\right)\left(1+f_{2}\right)
\label{eq:diffrate}
\end{eqnarray}
where $d^{3}\bar{p}_{V} = {d^{3}p_{V}/ 2E_{V}(2\pi)^{3}}$ and
similarly for the others, where the bar over the squared scattering amplitude
indicates that an initial spin average and final spin sum must been done,
where $f$ is again the Bose-Einstein distribution, the ${d}$\,s are degeneracy
factors, and the number density of $V$\,s is
\begin{equation}
n_{V} = d_{V} \int {d^{3}p_{V} \over (2\pi)^{3}} f_{V}.
\end{equation}
Reduction of phase space is accomplished with the aid of the four-momentum
conserving delta function and the remaining integration is done
numerically.  Processes considered in this study
are enumerated in Table~\ref{table:two} in no particular order and their
scattering amplitudes are relegated to the appendix.

\vskip 0.75 \baselineskip
\centerline{$\bbox{i.\,}$ $\bbox{\rho}$ \bf meson}
\vskip 0.75 \baselineskip

Elastic $\rho+\pi$ scattering proceeds through $s$ and $t$ channels
where $\pi, \omega, \phi, a_{1}$, and $\omega(1390)$ might be intermediate
or exchanged.  There are ten Feynman graphs in total.  Elastic scattering with
kaons may proceed through both $s$ and $t$ channels (two Feynman
graphs) with $K_{1}(1270)$ as the exchanged particle.  Though interference
effects are found to be quite modest the first ten graphs must be
added coherently upon squaring, i.e.  $|{\cal M}_{1}+{\cal M}_{2}+
\ldots+{\cal M}_{10}|^{2}$.  The result is then added to the coherent
sum of the two kaon graphs $|{\cal M}_{11}+{\cal M}_{12}|^{2}$.
In addition to these 12 graphs there is some possibility of
$\rho$+$\rho$ scattering.  It has been estimated to be relatively unimportant,
e.g. something like 5 MeV at 200 MeV temperature so it will not be discussed
further.  At the highest temperature considered the scattering rate
from the kaon contribution is 66 MeV while the sum of pion and kaon
contributions is 115.5 MeV.  Results are shown separately and then summed in
Fig.~\ref{fig:rhocoll}.  As expected,  the pions account for most of the
scattering throughout the entire temperature domain but at higher
temperatures the kaon contribution becomes nearly 50\% of the pions'.  At
highest temperature the collision rate is something like 75\% of the vacuum
decay rate.

\vskip 0.75 \baselineskip
\centerline{$\bbox{ii.\,}$ $\bbox{\omega}$ \bf meson}
\vskip 0.75 \baselineskip

The dominant contribution for $\omega$ scattering is from pions as well,
this time intermediate $\rho$ and $b_{1}(1235)$ mesons are important.  Four
graphs are needed to account for both $s$ and $t$ channels and must be
added coherently upon squaring $|{\cal M}_{1}+{\cal M}_{2}+\ldots+
{\cal M}_{4}|^{2}$.  Strange-particle processes involving the $\omega$ are
suppressed as compared with the $\rho$ since there is no resonance
playing a role as prominent as $K_{1}$.  But in their place a
sizable contribution comes from $\omega+\pi\to \pi+\pi$.
Two diagrams must be computed whose scattering amplitudes are listed along
with all the others in the appendix.  Resulting contributions are shown in
Fig.~\ref{fig:omecoll}.  A
fairly large rate is obtained at high temperature.  The unstable
$\omega$, whose vacuum width is about 8 MeV, will have an ``effective
width'' of 103 MeV assuming its decays are not drastically altered.  Recent
suggestions were made that a signature of dense hadronic matter might
be enhanced $e^{+}e^{-}$ production with invariant mass very near $m_{\omega}$
due to three pion initial states $\pi^{+}\pi^{-}\pi^{0}$ and owing to the
high densities\cite{pl94}.  A broadening to the $\omega$ of such an extent
as 95 MeV might significantly affect the possible signature.

\vskip 0.75 \baselineskip
\centerline{$\bbox{iii.\,}$ $\bbox{\phi}$ \bf meson}
\vskip 0.75 \baselineskip

No single channel dominates the $\phi$ like the other vector mesons
considered here since there is no strong $\phi + x$ resonance where $x$ might
be a pion, kaon, $K^{*}$, or others.  Many channels must therefore be
considered and are listed in table~\ref{table:two}.  For illustration
purposes the channel $\phi+\rho\to K_{1}(1270)+K$ is isolated and shown in
Fig.~\ref{fig:phirhorate}.  The solid curve results from using an energy
and temperature dependent width in the propagator for the exchanged kaon.
Notably, the rate is about 3 MeV at high temperatures.  To visualize the
sensitivity of the energy dependence of $\Gamma_{K}$, results are also
shown wherein fixed widths of 20 and 30 MeV are used instead of
Eq.~(\ref{eq:gammaofe}).
Clearly these energy dependences are quite important, especially for the
highest temperatures.  This channel is not an obvious candidate for a
large contribution to the rate yet turns out to be important.
As expected from simple number-density arguments, a large contribution
comes from the reaction $\phi+\pi\to K^{*}+K$.  Others are important
as can be seen from Fig.~\ref{fig:collcompare} which
compares contributions from a subset of the processes listed
in Table~\ref{table:two}.  The overall normalization of the
$\phi+\rho\to K_{1}(1270)+K$
reaction is somewhat of a surprise.  Other small surprises are
the strengths of $\phi+\pi \to \pi+\omega$
(long-dashed curve), $\phi+K^{*}\to \pi+K$ (lower-dotted curve)
and $\phi+K\to \pi + K^{*}$ (quadruple-dot-dashed curve).
At high temperatures several channels contribute nearly equally strongly
as can be seen from the column of $T$ = 200 MeV results shown numerically
in Table~\ref{table:two}.
Although quite small, contribution from the $\rho(1450)$ channel
(triple-dot-dashed curve) is also shown in the figure.

Then, the sum of all contributions is shown in Fig.~\ref{fig:totalrate}.
At 100 MeV temperature the rate is only 2.4 MeV.  Modification of the
$\phi$ width due to the matter is modest.  But as the temperature rises to
150 MeV the scattering rate is 8.6 MeV and at $T=$ 200 MeV, the rate goes
to 27.4 MeV.  Modification of this extent for a particle whose vacuum decay
width is of order 4 MeV is quite significant.

\subsection{Mean Free Paths}

The ``rule of thumb'' for hadronic collsion distances in finite temperature
media are of order 1 fm.  For $\rho$ and possibly $\omega$ mesons this is
reasonable since they react vigorously with pions.  However,
standard lore has it that only a small fraction of $\phi$ mesons in
these media will decay or interact before leaving.  To test this
idea for how frequently they interact, their mean free paths must be computed.
They can be approximated by
\begin{equation}
\overline{\lambda}_{V} = \bar{v}_{V}\, \left/ \, \,
\overline{\Gamma}^{\, \rm coll}_{V} \right. .
\end{equation}
Collision rates are already obtained, what remains to compute are
average velocities.  They are estimated with
\begin{equation}
\bar{v}_{V} = {d_{V}\over n_{V}} \int {d^{3}p_{V} \over
(2\pi)^{3}} f_{V} {|\,\bbox{p_{V}}|\over E_{V}}
\end{equation}
to be monotonically increasing functions of temperature.  Values for
the $\phi$ of $\bar{v}_{\phi}$ = 0.46, 0.54 and 0.61$\times c$ are
obtained
at temperatures 100, 150 and 200 MeV, respectively.  This translates into
mean free paths for the $\phi$ of 38.2, 14.9 and 4.4 fm at the corresponding
temperatures.  Results for $\rho$, $\omega$ and $\phi$ and for arbitrary
temperatures 100 $\le T \le$ 200 MeV are shown in Fig~\ref{fig:mfp}.  The
higher-temperature results for the $\phi$ are not consistent with common
lore and are therefore worth restating in more direct terms.  The mean free
path $\overline{\lambda}_{\phi}$ is shorter than a typical transverse size for
colliding nuclei which means on average, the $\phi$ will indeed scatter
before leaving the reaction zone.   Consequently, the assumption made earlier
about its momentum distribution being thermal is probably not unreasonable.

\section{Summary}
\label{sec:summary}

In this paper results were reported for collision rates of $\rho$,
$\omega$ and $\phi$ mesons at nonzero temperature.  Special attention was
given to modification of the zero-temperature scalar-boson propagator due to
the presence of matter since it is involved quite crucially in certain
reactions.  The real part of this modification was ignored while the
imaginary part was computed using a simple expression from kinetic theory
and then inserted by hand.

At low temperatures, by which 100 MeV or so is meant, the collision rates
for all three were quite small $\sim$ 10, 6, and 2 MeV for $\rho$,
$\omega$ and $\phi$.  Results for each of them rises monotonicaly
as the number densities of pions and kaons increase.  As temperatures
rise above the pion mass, the number densities of species other than pions
and kaons become large enough so that collisions with the $\phi$ are more
frequent.  Scattering of $\phi$ mesons with $\rho$\ and $K^{*}$ mesons
becomes significant around 150 MeV temperature while at high temperatures
several processes contribute nearly equally resulting in a total rate of 27.2
MeV at $T=200$ MeV.  At this extreme, the mean free path of $\phi$ mesons
is 4.4 fm!  One could be justified in using thermal distributions for this
species\cite{remark}.
These time and distance scales could have noticeable consequences on some
observables at least for $\omega$ and $\phi$-related ones.  A place to look
for modification might be in dilepton invariant mass spectra.  For
instance, the $\phi$ peak might appear rather differently and may not be so
clearly visible above the tail of $\pi^{+}\pi^{-}$ annihilations.  This
broadening or ``melting'' of the resonances could have an affect on the
overall normalization of low-mass $e^{+}e^{-}$ production since it depends
on a sum over all scattering energies for the hadrons (pions).  The broad
widths will change the normalization by roughly
$(\Gamma_{\rm narrow}/\Gamma_{\rm broad})^{2}$.  In the cases of
$\rho$ and $\omega$ at highest temperature these factors are 1/3 and 1/150!

A final remark is perhaps in order about the limitations of this approach to
studying collisional phenomena as compared with that of purely field theory.
In field theory at finite temperature the vector meson self-energy
approximated with one-loop diagrams gives rise to on-shell decays
within matter.  This is so because there is a basic connection
between the statistical mechanical collision rate $\Gamma$ and the imaginary
part of the field-theoretic finite-temperature self-energy $\Pi$\cite{weldon2}.
The two-loop contribution to $\Pi$ acquires an imaginary part at
$T\ne$ 0 due to scattering plus higher-order vertex corrections and
phase-space broadening of the final states in the above mentioned decays.
Mass modification due to the
matter is also present in such calculations since the real parts are
included.  Kinetic theory calculations, on the other hand, like the one
presented here simply cannot do all this.  Computing the self-energy to
two-loop order at finite-temperature requires much more effort than invested
in this work but should be done for a consistency check.

\section*{Acknowledgements}

It is a pleasure to thank S. Pratt for valuable discussions.
This work was supported by the National Science Foundation under grant number
PHY-9403666.

\appendix
\section*{Amplitudes}

Scattering amplitudes are included here for completeness.  The processes to
which they apply are enumerated in Table~\ref{table:two} of the text.  Many
of them contain the form factor $F(t)$ which is Eq.~(\ref{eq:ffactor}) in the
text and the first two $\rho$-meson amplitudes contain the form factor
$\tilde{F}_{\pi}(t)$ of Eq.~(\ref{eq:fgactor}).  Every external spin-one
particle in a Feynman graph introduces a polarization four-vector
$\epsilon^{\mu}$.  It depends on some momentum $\bbox{p}$ as well as a spin
index $\lambda$.  In the amplitudes written here, the spin indices are all
suppressed.  Upon squaring these amplitudes spin sums must be performed, giving
\begin{equation}
\sum\limits_{\lambda} \epsilon^{(\lambda)}_{\mu}(\bbox{p})\epsilon^{(\lambda)\,
*}_{\nu}(\bbox{p}) = -\left(g_{\mu\nu} - p_{\mu}p_{\nu}/m^{2}\right).
\end{equation}

\vskip 0.75 \baselineskip
\centerline{$\bbox{i.\,}$ $\bbox{\rho}$ \bf scattering amplitudes}
\vskip 0.75 \baselineskip

The individual amplitudes are
\begin{eqnarray}
{\cal M}_{1}^{(\rho)} &=& g_{\rho\pi\pi}^{2}\, \epsilon^{\mu}
(p_{\rho})\left(2p_{\pi}+p_{\rho}\right)_{\mu}\, \epsilon^{\nu}
(p_{\rho^{\prime}})
\left(2p_{\rho^{\prime}}+p_{\pi^{\prime}}\right)_{\nu}\,
\Delta_{\pi}(s, E, T) \nonumber\\
{\cal M}_{2}^{(\rho)} &=& g_{\rho\pi\pi}^{2}\, \tilde{F}_{\pi}^{2}(t)\,
\epsilon^{\mu}
(p_{\rho})\left(2p_{\pi^{\prime}}-p_{\rho}\right)_{\mu}\, \epsilon^{\nu}
(p_{\rho^{\prime}})
\nonumber\\
& & \quad\times \left(2p_{\pi}-p_{\rho^{\prime}}\right)_{\nu}\,
\Delta_{\pi}(l^{2},E,T) \nonumber\\
{\cal M}_{3}^{(\rho)} &=& g_{\rho\pi\omega}^{2}\, \epsilon_{\mu\nu\alpha\beta}
\, p_{\rho}^{\mu}\, \epsilon^{\nu}(p_{\rho})q^{\alpha}
\left(g^{\beta\tau}-q^{\beta}q^{\tau}/m_{\omega}^{2}\right)q^{\sigma}
\epsilon^{\lambda}(p_{\rho^{\prime}})
\nonumber\\
& & \quad\times p_{\rho^{\prime}}^{\kappa}\, \epsilon_{\kappa\lambda\sigma\tau}
{1\over s-m_{\omega}^{2}-im_{\omega}\Gamma_{\omega}} \nonumber\\
{\cal M}_{4}^{(\rho)} &=& g_{\rho\pi\omega}^{2}\, F^{2}_{\omega}(t)\,
\epsilon_{\mu\nu\alpha\beta}
\, p_{\rho}^{\mu}\, \epsilon^{\nu}(p_{\rho})
l^{\alpha}\left(g^{\beta\tau}-
l^{\beta}l^{\tau}/m_{\omega}^{2}\right)l^{\sigma}
\epsilon^{\lambda}(p_{\rho^{\prime}})
\nonumber\\
& & \quad\times p_{\rho^{\prime}}^{\kappa}\, \epsilon_{\kappa\lambda\sigma\tau}
{1\over l^{2}-m_{\omega}^{2}-im_{\omega}\Gamma_{\omega}} \nonumber\\
{\cal M}_{5}^{(\rho)} &=& g_{\rho\pi\phi}^{2}\, \epsilon_{\mu\nu\alpha\beta}
\, p_{\rho}^{\mu}\, \epsilon^{\nu}(p_{\rho})q^{\alpha}
\left(q^{\beta\tau}-q^{\beta}q^{\tau}/m_{\phi}^{2}\right)q^{\sigma}
\epsilon^{\lambda}(p_{\rho^{\prime}})
\nonumber\\
& & \quad\times p_{\rho^{\prime}}^{\kappa}\, \epsilon_{\kappa\lambda\sigma\tau}
{1\over s-m_{\phi}^{2}-im_{\phi}\Gamma_{\phi}} \nonumber\\
{\cal M}_{6}^{(\rho)} &=& g_{\rho\pi\phi}^{2}\, F^{2}_{\phi}(t)\,
\epsilon_{\mu\nu\alpha\beta}
\, p_{\rho}^{\mu}\, \epsilon^{\nu}(p_{\rho})l^{\alpha}
\left(q^{\beta\tau}-l^{\beta}l^{\tau}/m_{\phi}^{2}\right)l^{\sigma}
\epsilon^{\lambda}(p_{\rho^{\prime}})
\nonumber\\
& & \quad\times p_{\rho^{\prime}}^{\kappa}\, \epsilon_{\kappa\lambda\sigma\tau}
{1\over l^{2}-m_{\phi}^{2}-im_{\phi}\Gamma_{\phi}} \nonumber\\
{\cal M}_{7}^{(\rho)} &=& g_{\rho a_{1}\pi}^{2} \epsilon_{\sigma}
(p_{\rho})\, \left[p_{\rho}^{\mu}g_{\nu\sigma}-p_{\rho}^{\nu}
g_{\mu\sigma}\right] \left[q_{\mu}g_{\nu\sigma}
-q_{\nu}g_{\mu\alpha}
\right] \nonumber\\
& & \quad\times \left[g^{\alpha\beta}-q^{\alpha}
q^{\beta}/m_{a_{1}}^{2}\right]
\left[q_{\kappa}g_{\lambda\beta}-q_{\lambda}
g_{\kappa\beta}\right] \nonumber\\
& & \quad\times
\left[p_{\rho^{\prime}}^{\kappa}g^{\lambda\tau}-p_{\rho^{\prime}}^{\lambda}
g^{\kappa\tau}
\right] \epsilon_{\tau}(p_{\rho^{\prime}})
{1\over s-m_{a_{1}}^{2}-im_{a_{1}}
\Gamma_{a_{1}}} \nonumber\\
{\cal M}_{8}^{(\rho)} &=& g_{\rho a_{1}\pi}^{2}\, F^{2}_{a_{1}}(t)\,
\epsilon_{\sigma}
(p_{\rho})\, \left[p_{\rho}^{\mu}g_{\nu\sigma}-p_{\rho}^{\nu}
g_{\mu\sigma}\right] \left[l_{\mu}g_{\nu\sigma}
-l{\nu}g_{\mu\alpha}
\right] \nonumber\\
& & \quad\times \left[g^{\alpha\beta}-l^{\alpha}
l^{\beta}/m_{a_{1}}^{2}\right]
\left[l_{\kappa}g_{\lambda\beta}-
l_{\lambda}g_{\kappa\beta}\right] \nonumber\\
& & \quad\times
\left[p_{\rho^{\prime}}^{\kappa}g^{\lambda\tau}-p_{\rho^{\prime}}^{\lambda}
g^{\kappa\tau}
\right] \epsilon_{\tau}(p_{\rho^{\prime}})
{1\over l^{2}-m_{a_{1}}^{2}-im_{a_{1}}\Gamma_{a_{1}}} \nonumber\\
{\cal M}_{9}^{(\rho)} &=& g_{\rho\pi\omega^{*}}^{2}\,
\epsilon_{\mu\nu\alpha\beta}\, p_{\rho}^{\mu}\,
\epsilon^{\nu}(p_{\rho})\, q^{\alpha}
\left(g^{\beta\tau}-q^{\beta}q^{\tau}/m_{\omega^{*}}^{2}\right)
q^{\sigma}
\epsilon^{\lambda}(p_{\rho^{\prime}})
\nonumber\\
& & \quad\times p_{\rho^{\prime}}^{\kappa}\, \epsilon_{\kappa\lambda\sigma\tau}
{1\over s-m_{\omega^{*}}^{2}-im_{\omega^{*}}\Gamma_{\omega^{*}}} \nonumber\\
{\cal M}_{10}^{(\rho)} &=& g_{\rho\pi\omega^{*}}^{2}\, F^{2}_{\omega^{*}}(t)\,
\epsilon_{\mu\nu\alpha\beta}
\, p_{\rho}^{\mu}\, \epsilon^{\nu}(p_{\rho})
l^{\alpha}\left(g^{\beta\tau}-
l^{\beta}l^{\tau}/m_{\omega^{*}}^{2}\right)l^{\sigma}
\epsilon^{\lambda}(p_{\rho^{\prime}})
\nonumber\\
& & \quad\times p_{\rho^{\prime}}^{\kappa}\, \epsilon_{\kappa\lambda\sigma\tau}
{1\over l^{2}-m_{\omega^{*}}^{2}-im_{\omega^{*}}\Gamma_{\omega^{*}}}
\nonumber\\
{\cal M}_{11}^{(\rho)} &=& g_{\rho K_{1}K}^{2} \epsilon_{\sigma}
(p_{\rho})\, \left[p_{\rho}^{\mu}g_{\nu\sigma}-p_{\rho}^{\nu}
g_{\mu\sigma}\right] \left[q_{\mu}g_{\nu\sigma}
-q_{\nu}g_{\mu\alpha}
\right] \nonumber\\
& & \quad\times \left[g^{\alpha\beta}-q^{\alpha}
q^{\beta}/m_{K_{1}}^{2}\right]
\left[q_{\kappa}g_{\lambda\beta}-q_{\lambda}
g_{\kappa\beta}\right] \nonumber\\
& & \quad\times
\left[p_{\rho^{\prime}}^{\kappa}g^{\lambda\tau}-p_{\rho^{\prime}}^{\lambda}
g^{\kappa\tau}
\right] \epsilon_{\tau}(p_{\rho^{\prime}})
{1\over s-m_{K_{1}}^{2}-im_{K_{1}}
\Gamma_{K_{1}}} \nonumber\\
{\cal M}_{12}^{(\rho)} &=& g_{\rho K_{1}K}^{2}\, F^{2}_{K_{1}}(t)\,
\epsilon_{\sigma}
(p_{\rho})\, \left[p_{\rho}^{\mu}g_{\nu\sigma}-p_{\rho}^{\nu}
g_{\mu\sigma}\right] \left[\tilde{l}_{\mu}g_{\nu\sigma}
-\tilde{l}{\nu}g_{\mu\alpha}
\right] \nonumber\\
& & \quad\times \left[g^{\alpha\beta}-\tilde{l}^{\alpha}
\tilde{l}^{\beta}/m_{K_{1}}^{2}\right]
\left[\tilde{l}_{\kappa}g_{\lambda\beta}-
\tilde{l}_{\lambda}g_{\kappa\beta}\right] \nonumber\\
& & \quad\times
\left[p_{\rho^{\prime}}^{\kappa}g^{\lambda\tau}-p_{\rho^{\prime}}^{\lambda}
g^{\kappa\tau}
\right] \epsilon_{\tau}(p_{\rho^{\prime}})
{1\over \tilde{l}^{2}-m_{K_{1}}^{2}-im_{K_{1}}\Gamma_{K_{1}}} \nonumber\\
\end{eqnarray}
where the primed variables indicate final states, where $\omega^{*}$ is
shorthand for $\omega(1390)$, $q=p_{\rho}+p_{\pi}$ and naturally $s=q^{2}$,
$\tilde{q}=p_{\rho}+p_{K}$,
$l=p_{\pi^{\prime}}-p_{\rho}$ and $\tilde{l}=p_{K^{\prime}}-p_{\rho}$,
and where Eq.~(\ref{eq:prop}) has been utilized for compactness.

\vskip 0.75 \baselineskip
\centerline{$\bbox{ii.\,}$ $\bbox{\omega}$ \bf scattering amplitudes}
\vskip 0.75 \baselineskip

\begin{eqnarray}
{\cal M}_{1}^{(\omega)} &=& g_{\rho b_{1}\pi}^{2} \epsilon_{\sigma}
(p_{\omega})\, \left[p_{\omega}^{\mu}g_{\nu\sigma}-p_{\omega}^{\nu}
g_{\mu\sigma}\right] \left[q_{\mu}g_{\nu\sigma}
-q_{\nu}g_{\mu\alpha}
\right] \nonumber\\
& & \quad\times \left[g^{\alpha\beta}-q^{\alpha}
q^{\beta}/m_{b_{1}}^{2}\right]
\left[q_{\kappa}g_{\lambda\beta}-q_{\lambda}
g_{\kappa\beta}\right] \nonumber\\
& & \quad\times
\left[p_{\omega^{\prime}}^{\kappa}g^{\lambda\tau}-p_{\omega^{\prime}}^{\lambda}
g^{\kappa\tau}
\right] \epsilon_{\tau}(p_{\omega^{\prime}})
{1\over s-m_{b_{1}}^{2}-im_{b_{1}}
\Gamma_{b_{1}}} \nonumber\\
{\cal M}_{2}^{(\omega)} &=& g_{\rho b_{1}\pi}^{2}\, F^{2}_{b_{1}}(t)\,
\epsilon_{\sigma}
(p_{\omega})\, \left[p_{\omega}^{\mu}g_{\nu\sigma}-p_{\omega}^{\nu}
g_{\mu\sigma}\right] \left[l_{\mu}g_{\nu\sigma}
-l{\nu}g_{\mu\alpha}
\right] \nonumber\\
& & \quad\times \left[g^{\alpha\beta}-l^{\alpha}
l^{\beta}/m_{a_{1}}^{2}\right]
\left[l_{\kappa}g_{\lambda\beta}-
l_{\lambda}g_{\kappa\beta}\right] \nonumber\\
& & \quad\times
\left[p_{\omega^{\prime}}^{\kappa}g^{\lambda\tau}-p_{\omega^{\prime}}^{\lambda}
g^{\kappa\tau}
\right] \epsilon_{\tau}(p_{\omega^{\prime}})
{1\over l^{2}-m_{b_{1}}^{2}-im_{b_{1}}\Gamma_{b_{1}}} \nonumber\\
{\cal M}_{3}^{(\omega)} &=& g_{\omega\pi\rho}^{2}\,
\epsilon_{\mu\nu\alpha\beta}
\, p_{\omega}^{\mu}\, \epsilon^{\nu}(p_{\omega})q^{\alpha}
\left(g^{\beta\tau}-q^{\beta}q^{\tau}/m_{\rho}^{2}\right)q^{\sigma}
\epsilon^{\lambda}(p_{\omega^{\prime}})
\nonumber\\
& & \quad\times p_{\omega^{\prime}}^{\kappa}\,
\epsilon_{\kappa\lambda\sigma\tau}
{1\over s-m_{\rho}^{2}-im_{\rho}\Gamma_{\rho}} \nonumber\\
{\cal M}_{4}^{(\omega)} &=& g_{\omega\pi\rho}^{2}\, F^{2}_{\rho}(t)\,
\epsilon_{\mu\nu\alpha\beta}
\, p_{\omega}^{\mu}\, \epsilon^{\nu}(p_{\omega})
l^{\alpha}\left(g^{\beta\tau}-
l^{\beta}l^{\tau}/m_{\rho}^{2}\right)l^{\sigma}
\epsilon^{\lambda}(p_{\omega^{\prime}})
\nonumber\\
& & \quad\times p_{\omega^{\prime}}^{\kappa}\,
\epsilon_{\kappa\lambda\sigma\tau}
{1\over l^{2}-m_{\rho}^{2}-im_{\rho}\Gamma_{\rho}} \nonumber\\
{\cal M}_{5}^{(\omega)} &=& g_{\omega\pi\rho}\,g_{\rho\pi\pi}\,
\epsilon_{\mu\nu\alpha\beta}\, p_{\omega}^{\mu}\epsilon^{\nu}(p_{\omega})
q^{\alpha} \left(g^{\beta\lambda}-q^{\beta}q^{\lambda}/m_{\rho}^{2}\right)
(p_{\pi^{\prime}}-p_{\pi^{\prime\prime}})_{\lambda} \nonumber\\
& & \quad\times
{1\over s-m_{\rho}^{2}-im_{\rho}\Gamma_{\rho}} \nonumber\\
{\cal M}_{6}^{(\omega)} &=& g_{\omega\pi\rho}\,g_{\rho\pi\pi}\, F^{2}_{\rho}(t)
\,\epsilon_{\mu\nu\alpha\beta}\, p_{\omega}^{\mu}\epsilon^{\nu}(p_{\omega})
l^{\alpha} \left(g^{\beta\lambda}-l^{\beta}l^{\lambda}/m_{\rho}^{2}\right)
\nonumber\\
& & \quad\times (p_{\pi}+p_{\pi^{\prime}})_{\lambda}
{1\over l^{2}-m_{\rho}^{2}-im_{\rho}\Gamma_{\rho}} \nonumber\\
\end{eqnarray}

where this time $q=p_{\omega}+p_{\pi}$ and $l=p_{\pi^{\prime}}-p_{\omega}$.

\vskip 0.75 \baselineskip
\centerline{$\bbox{ii.\,}$ $\bbox{\phi}$ \bf scattering amplitudes}
\vskip 0.75 \baselineskip

The scattering amplitudes for $\phi$ reactions are

\begin{eqnarray}
{\cal M}_{1}^{(\phi)} &=& g_{\phi KK}\, g_{K_{1}\rho K}\,
F_{K}^{2}(t) \epsilon^{\kappa}
(p_{\phi})\left(2p_{K}-p_{\phi}\right)_{\kappa}\, \epsilon^{\sigma}(p_{K_{1}})
\nonumber\\
& & \quad\times
\left[p_{K_{1}\, \mu}g_{\sigma\nu}-p_{K_{1}\, \nu}g_{\sigma\mu}
\right]
\left[p_{\rho}^{\mu}g^{\lambda\nu}-p_{\rho}^{\nu}g^{\lambda\mu}
\right]
\epsilon_{\lambda}(p_{\rho}) \Delta_{K}(\tilde{l}^{2},E,T)
\nonumber\\
{\cal M}_{2}^{(\phi)} &=& g_{\phi KK}\, g_{K^{*}\pi K}\, F_{K}^{2}(t)\,
\epsilon^{\mu}
(p_{\phi})\left(2p_{K}-p_{\phi}\right)_{\mu}\, \epsilon^{\nu}(p_{K^{*}})
\left(2p_{\pi}-p_{K^{*}}\right)_{\nu}
\nonumber\\ & & \quad\times \Delta_{K}(\tilde{l}^{2},E,T)
\nonumber\\
{\cal M}_{3}^{(\phi)} &=& g_{\phi KK}\, g_{K^{*}\pi K}\, F_{K}^{2}(t)\,
\epsilon^{\mu}(p_{\phi})\left(2p_{K}-p_{\phi}\right)_{\mu}\,
\epsilon^{\nu}(p_{K^{*}})
\left(2p_{\pi}-p_{K^{*}}\right)_{\nu}
\nonumber\\ & & \quad\times \Delta_{K}(\tilde{l}^{2},E,T)
\nonumber\\
{\cal M}_{4}^{(\phi)} &=& g_{\phi KK}\, g_{K_{1}\rho K}\, F_{K}^{2}(t)\,
\epsilon^{\kappa}
(p_{\phi})\left(2p_{K}-p_{\phi}\right)_{\kappa}\, \epsilon^{\sigma}(p_{K_{1}})
\nonumber\\
& & \quad\times
\left[p_{K_{1}\, \mu}g_{\sigma\nu}-p_{K_{1}\, \nu}g_{\sigma\mu}
\right]
\left[p_{\rho}^{\mu}g^{\lambda\nu}-p_{\rho}^{\nu}g^{\lambda\mu}
\right]
\epsilon_{\lambda}(p_{\rho}) \Delta_{K}(\tilde{l}^{2},E,T)
\nonumber\\
{\cal M}_{5}^{(\phi)} &=& g_{\phi\rho\pi}\, g_{\rho\pi\pi}\, F_{\pi}^{2}(t)\,
\epsilon_{\mu\nu\alpha\beta}\, p_{\phi}^{\mu} \, \epsilon^{\nu}(p_{\phi})
\, p_{\rho}^{\alpha}\, \epsilon^{\beta}(p_{\rho})
\epsilon^{\lambda}(p_{\rho})\left(2p_{\pi}-p_{\rho}\right)_{\lambda}
\Delta_{\pi}(l^{2},E,T)
\nonumber\\
{\cal M}_{6}^{(\phi)} &=& g_{\phi\rho\pi}\, g_{a_{1}\rho\pi}\, F_{\pi}^{2}(t)\,
\epsilon_{\mu\nu\alpha\beta}\, p_{\phi}^{\mu}\, \epsilon^{\nu}\,
(p_{\phi})\, p_{\rho}^{\alpha}\epsilon^{\beta}
(p_{\rho})\, \epsilon^{\sigma}(p_{a_{1}})
\nonumber\\
& & \quad\times
\left[p_{a_{1}\, \tau}g_{\sigma\kappa}-p_{a_{1}\, \kappa}g_{\sigma\tau}
\right]
\left[p_{\rho}^{\tau}g^{\lambda\kappa}-p_{\rho}^{\kappa}g^{\lambda\tau}
\right]
\epsilon_{\lambda}(p_{\rho}) \Delta_{\pi}(l^{2},E,T)
\nonumber\\
{\cal M}_{7}^{(\phi)} &=& g_{\phi b_{1}\pi}\,g_{\omega b_{1}\pi}
\epsilon_{\sigma}
(p_{\phi})\, \left[p_{\phi}^{\mu}g_{\nu\sigma}-p_{\phi}^{\nu}
g_{\mu\sigma}\right] \left[q_{\mu}g_{\nu\sigma}
-q_{\nu}g_{\mu\alpha}
\right] \nonumber\\
& & \quad\times \left[g^{\alpha\beta}-q^{\alpha}
q^{\beta}/m_{a_{1}}^{2}\right]
\left[q_{\kappa}g_{\lambda\beta}-q_{\lambda}
g_{\kappa\beta}\right] \nonumber\\
& & \quad\times
\left[p_{\omega}^{\kappa}g^{\lambda\tau}-p_{\omega}^{\lambda}
g^{\kappa\tau}
\right] \epsilon_{\tau}(p_{\omega})
{1\over s-m_{b_{1}}^{2}-im_{b_{1}}
\Gamma_{b_{1}}} \nonumber\\
{\cal M}_{8}^{(\phi)} &=& g_{\phi b_{1}\pi}\,g_{\omega b_{1}\pi}
F^{2}_{b_{1}}(t)\, \epsilon_{\sigma}
(p_{\phi})\, \left[p_{\phi}^{\mu}g_{\nu\sigma}-p_{\phi}^{\nu}
g_{\mu\sigma}\right] \left[l_{\mu}g_{\nu\sigma}
-l{\nu}g_{\mu\alpha}
\right] \nonumber\\
& & \quad\times \left[g^{\alpha\beta}-l^{\alpha}
l^{\beta}/m_{b_{1}}^{2}\right]
\left[l_{\kappa}g_{\lambda\beta}-
l_{\lambda}g_{\kappa\beta}\right] \nonumber\\
& & \quad\times
\left[p_{\omega}^{\kappa}g^{\lambda\tau}-p_{\omega}^{\lambda}
g^{\kappa\tau}
\right] \epsilon_{\tau}(p_{\rho^{\prime}})
{1\over l^{2}-m_{b_{1}}^{2}-im_{b_{1}} \Gamma_{b_{1}}} \nonumber\\
%
{\cal M}_{9}^{(\phi)} &=& g_{\phi KK}^{2}\,
\epsilon^{\mu}(p_{\phi})\left(2p_{K}+p_{\phi}\right)_{\mu}\,
\epsilon^{\nu}(p_{\phi^{\prime}})\left(2p_{K^{\prime}}+p_{\phi^{\prime}}
\right)_{\nu}
\nonumber\\ & & \quad\times {1 \over s-m_{K}^{2}-im_{K}\Gamma_{K}(E,T)}
\nonumber\\
{\cal M}_{10}^{(\phi)} &=& g_{\phi KK}^{2}\, F_{K}^{2}(t) \,
\epsilon^{\mu}(p_{\phi})\left(2p_{K^{\prime}}-p_{\phi}\right)_{\mu}\,
\epsilon^{\nu}(p_{\phi^{\prime}})\left(2p_{K^{\prime}}-p_{\phi^{\prime}}
\right)_{\nu}
\nonumber\\ & & \quad\times \Delta_{K}(\tilde{l}^{2},E,T)
\nonumber\\
{\cal M}_{11}^{(\phi)} &=& g_{\phi KK}\, g_{\rho KK}\, F_{K}^{2}(t) \,
\epsilon^{\mu}
(p_{\phi})\left(2p_{K^{\prime}}-p_{\phi}\right)_{\mu}\,
\epsilon^{\nu}(p_{\rho})
\left(2p_{K}-p_{\rho}\right)_{\nu}
\nonumber\\ & & \quad\times \Delta_{K}(\tilde{l}^{2},E,T)
\nonumber\\
{\cal M}_{12}^{(\phi)} &=& g_{\phi KK}^{2}\, F_{K}^{2}(t)\, \epsilon^{\mu}
(p_{\phi}) \left(2p_{K}-p_{\phi}\right)_{\mu}\,
\epsilon^{\nu}(p_{\tilde\phi})
\left(2p_{K^{\prime}}-p_{\tilde\phi}\right)_{\nu}
\nonumber\\ & & \quad\times \Delta_{K}(\tilde{l}^{2},E,T)
\nonumber\\
{\cal M}_{13}^{(\phi)} &=& g_{\phi\rho\pi}\, g_{K^{*}\pi K}\, F_{\pi}^{2}(t)\,
\epsilon_{\mu\nu\alpha\beta}\, p_{\phi}^{\mu}\, \epsilon^{\nu}
(p_{\phi}) \, p_{\rho}^{\alpha} \, \epsilon^{\beta}(p_{\rho})
\epsilon^{\lambda}(p_{K^{*}})
\left(2p_{K}-p_{K^{*}}\right)_{\lambda}
\nonumber\\ & & \quad\times \Delta_{\pi}(l^{2},E,T)
\nonumber\\
{\cal M}_{14}^{(\phi)} &=& g_{\phi\rho\pi}\, g_{K^{*}\pi K}\, F_{\pi}^{2}(t)\,
\epsilon_{\mu\nu\alpha\beta}\, p_{\phi}^{\mu}\, \epsilon^{\nu}
(p_{\phi}) \, p_{\rho}^{\alpha}\, \epsilon^{\beta}(p_{\rho})
\epsilon^{\lambda}(p_{K^{*}})\left(2p_{K}-p_{K^{*}}\right)_{\lambda}
\nonumber\\ & & \quad\times \Delta_{\pi}(l^{2},E,T)
\nonumber\\
{\cal M}_{15}^{(\phi)} &=& g_{\phi KK}\, g_{\rho KK}\, F_{\pi}^{2}(t)\,
\epsilon^{\mu}(p_{\phi})\, \left(2p_{K^{\prime}}-p_{\phi}\right)_{\mu}
\epsilon^{\nu}(p_{\rho})\, \left(2p_{K}-p_{\rho}\right)_{\nu}
\Delta_{K}(\tilde{l}^{2},E,T)
\nonumber\\
{\cal M}_{16}^{(\phi)} &=& g_{\phi\rho^{*}\pi}\, g_{\rho^{*}\pi\pi}
\, \epsilon_{\mu\nu\alpha\beta}\, p_{\phi}^{\mu}\, \epsilon^{\nu}(p_{\phi})\,
\left(p_{\phi}+p_{\pi}\right)^{\alpha}\left[g^{\beta\lambda}-
\left(p_{\phi}+p_{\pi} \right)^{\beta}\left(p_{\phi}+p_{\pi} \right)^{\lambda}
\right] \nonumber\\
& & \quad\times (p_{\pi^{\prime}}-p_{\pi^{\prime\prime}})_{\lambda}
{1\over s-m_{\rho^{*}}^{2}-im_{\rho^{*}}
\Gamma_{\rho^{*}}} \nonumber\\
{\cal M}_{17}^{(\phi)} &=& g_{\phi KK}\, g_{K^{*}\pi K}\,
\epsilon^{\mu}(p_{\phi})\left(p_{\pi}+p_{K}\right)_{\mu}
\epsilon^{\nu}(p_{K^{*}})
\left(p_{\pi}+p_{K}\right)_{\nu}
\nonumber\\ & & \quad\times {1 \over s-m_{K}^{2}-im_{K}\Gamma_{K}(E,T)}
\end{eqnarray}
where $\rho^{*}$ is shorthand for $\rho(1450)$,
the tildes in ${\cal M}_{12}$ are needed in order to tell the
two initial $\phi$\,s apart,
$q=p_{\phi}+p_{\pi}$, $l=p_{\pi^{\prime}}-p_{\omega}$,
$\tilde{l}=p_{K^{\prime}}-p_{\phi}$
and the primed and double-primed variables indicate and distinquish
between final states.

\begin{figure}
\caption{Scattering of the vector $V$ with hadron $\{a\}$ through
(a) an $s$-channel ``resonance'' $R$, and (b) through a $t$-channel
in which $E$ indicates the exchanged hadron.  Particles $\{a, 1, 2, R, E\}$
are enumerated in the text.}
\label{fig:diagram}
\end{figure}
\begin{figure}
\caption{Scattering rate or width for pions (a) and kaons (b) in
an ensemble of pions, kaons and $\rho$ mesons.}
\label{fig:piandkawidth}
\end{figure}
\begin{figure}
\caption{Collision rate of $\rho$ with pions and with kaons.}
\label{fig:rhocoll}
\end{figure}
\begin{figure}
\caption{Collision rate of $\omega$ with pions.}
\label{fig:omecoll}
\end{figure}
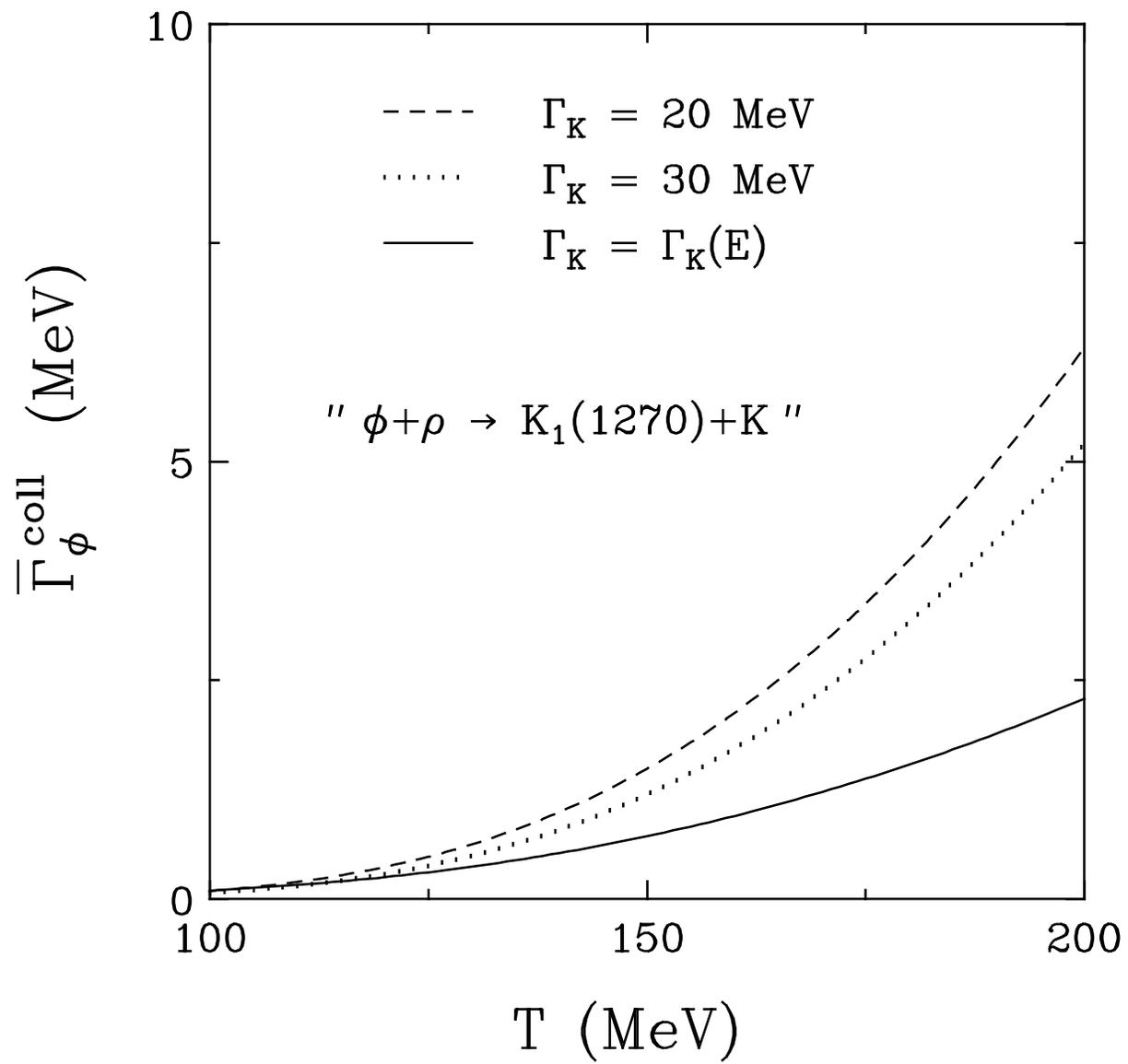
\begin{figure}
\caption{Collision rate of $\phi$ with $\rho$ mesons into a
$K_{1}(1270)+K$ final state.  Three different widths are used for the
exchanged kaon:  constant values of 20 and 30 MeV are shown as dashed and
dotted curves and an energy-dependent width results in the solid curve.}
\label{fig:phirhorate}
\end{figure}
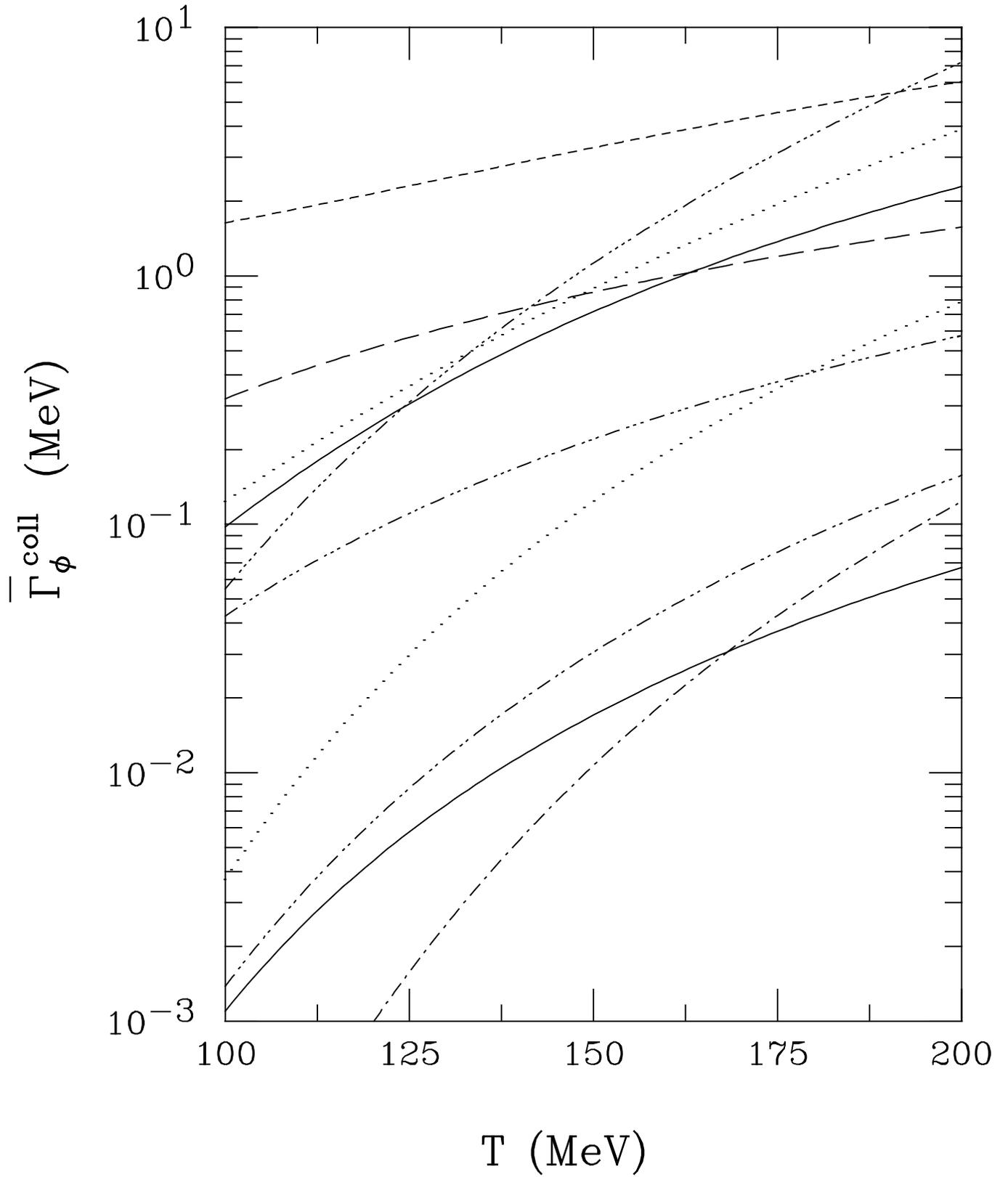
\begin{figure}
\caption{Contribution to the overall rate from various
channels. The curves correspond to the following reactions.  Upper solid is
$\phi+\rho\to K_{1}(1270)+K$, lower solid is $\phi+ \rho \to a_{1}(1260)+
\rho$, short-dashed is $\phi+\pi\to K^{*}+K$, long-dashed is $\phi + \pi
\to b_{1}(1235)\to \pi+\omega$,  upper-dotted is  $\phi+ K \to \phi + K$,
lower-dotted is $\phi+K^{*}\to \pi+K$, dot-dashed is $\phi+K_{1}(1270)\to
\rho + K$, double-dot-dashed refers to the reaction $\phi+\rho\to
\pi+\rho$, triple-dot-dashed refers to the reaction $\phi+\pi\to
\rho(1450)\to\pi+\pi$ and finally, quadruple-dot-dashed refers to
$\phi+K\to \pi+K^{*}$.}
\label{fig:collcompare}
\end{figure}
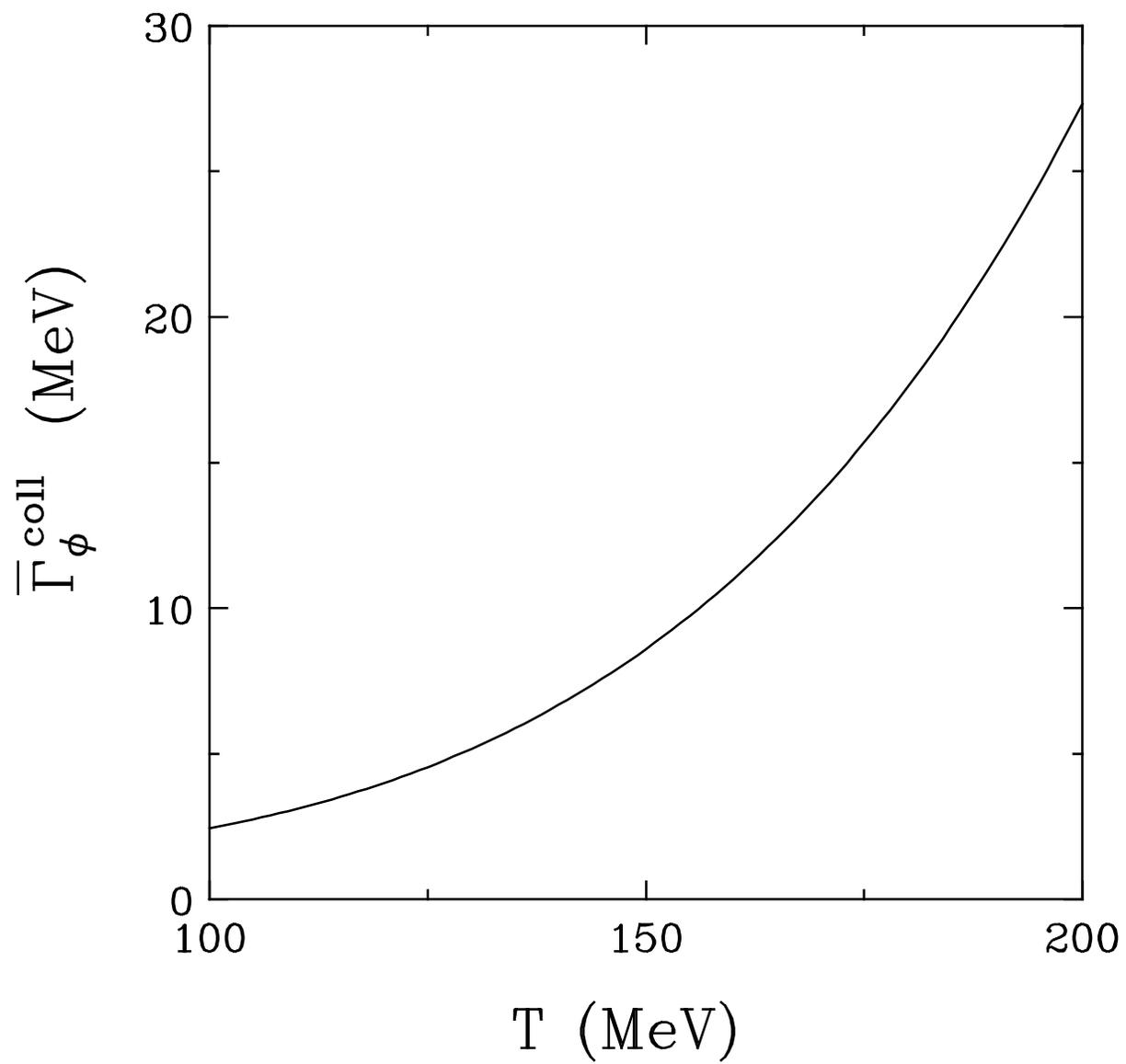
\begin{figure}
\caption{Total collision rate of $\phi$ mesons in hot hadronic
matter.}
\label{fig:totalrate}
\end{figure}
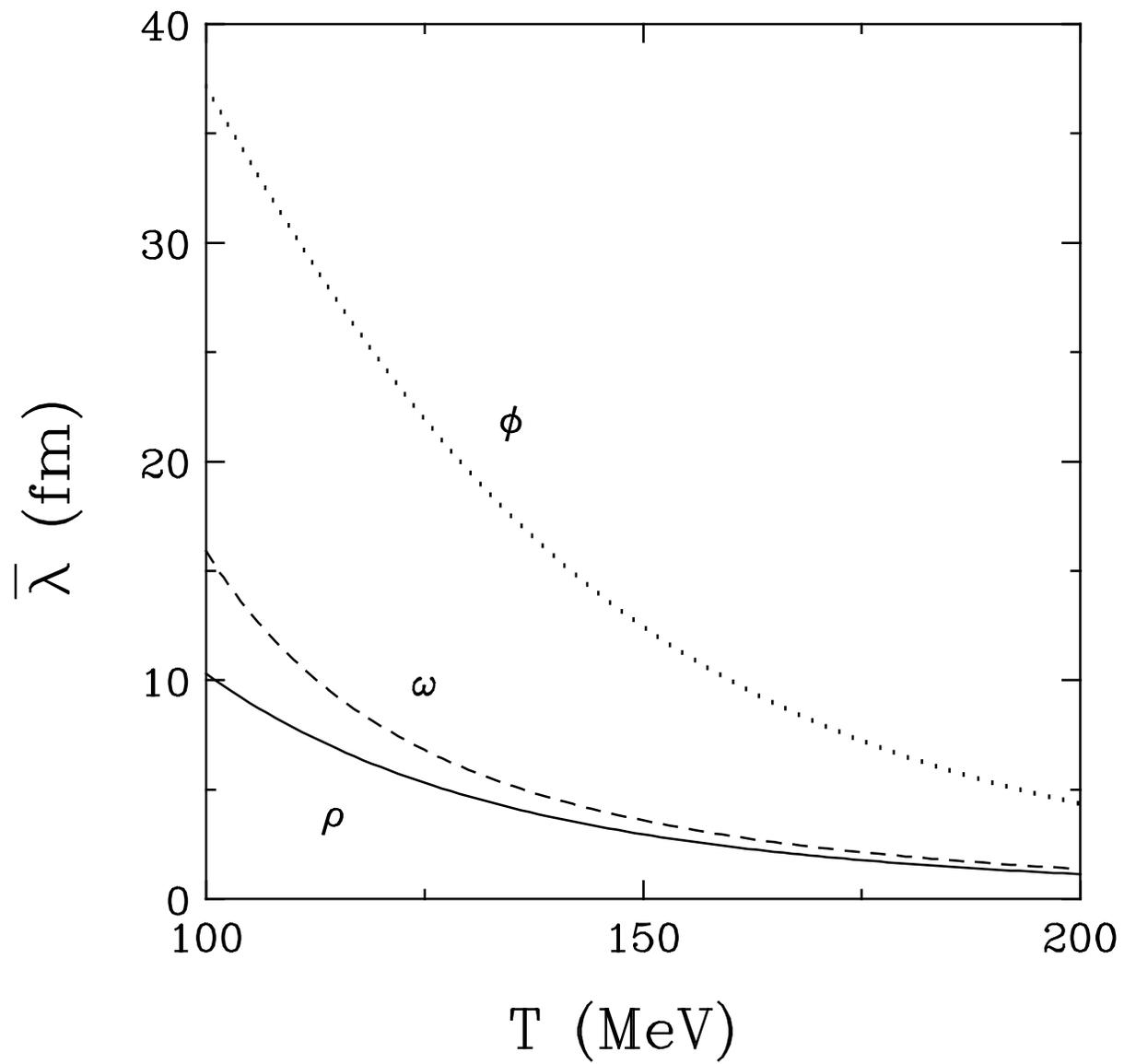
\begin{figure}
\caption{Mean free paths for $\rho$, $\omega$ and $\phi$ mesons in hot
hadronic matter.}
\label{fig:mfp}
\end{figure}
\narrowtext
\begin{table}
\caption{Coupling constants obtained from the decay widths in
Eq.~(\protect\ref{eq:elemdecay}).}
\vskip 0.05 true in
\begin{tabular}{cc}
coupling constant & numerical value \\
\tableline
$g_{\phi KK}^{2}/4\pi$  &  2.71 \\
$g_{\rho\pi\pi}^{2}/4\pi$  & 2.94 \\
$g_{K^{*} \pi K}^{2}/4\pi$  &  2.47 \\
$g_{K_{1} \rho K}^{2}/4\pi$  &  0.45 [GeV$^{-1}$] \\
$g_{\phi \rho\pi}^{2}/4\pi$  &  0.28 [GeV$^{-1}$] \\
$g_{a_{1} \rho \pi}^{2}/4\pi$  &  0.77 [GeV$^{-1}$] \\
$g_{b_{1} \omega \pi}^{2}/4\pi$  &  0.32 [GeV$^{-1}$] \\
$g_{b_{1} \phi \pi}^{2}/4\pi$  &  $7.6\times 10^{-3}$ [GeV$^{-1}$] \\
$g_{\rho KK}^{2}/4\pi$  & 0.73\footnote{This is taken from quark-model
arguments to be half of the $\rho\pi\pi$ coupling constant just as in
Ref.~\cite{seko}.} \\
$g_{\rho^{\prime}\phi\pi}^{2}/4\pi$  & 0.20 [GeV$^{-1}$] \\
$g_{\rho^{\prime}\pi\pi}^{2}/4\pi$  & 2.41\footnote{A branching fraction
of 90\% was assumed.} \\
$g_{\omega\pi\rho}^{2}/4\pi$  & 15.80\footnote{This is obtained from
vector-meson-dominance arguments to be $g_{\omega\pi\rho}=
(e/g_{\rho\pi\pi})^{-1}g_{\omega\pi\gamma}$.} \\
$g_{\rho\pi\omega^{\prime}}^{2}/4\pi$  & 6.53 [GeV$^{-1}$] \\
\end{tabular}
\label{table:one}
\end{table}
\mediumtext
\begin{table}
\caption{Reactions considered in which the vector meson scatters.}
\vskip 0.1 true in
\begin{tabular}{ccccc}
graph number & reaction & intermediate meson &
channels & $\overline{\Gamma}$($T$=200 MeV) \\
\tableline
\\
\multicolumn{5}{c}{$\rho$ meson reactions} \\
1, 2 & $\rho+\pi\to \rho+\pi$ & $\pi$ & $s$, $t$ &  \\
3, 4 & $\rho+\pi\to \rho+\pi$ & $\omega$ & $s$, $t$ & \\
5, 6 & $\rho+\pi\to \rho+\pi$ & $\phi$ &
{\ifnum\precount > \number\zero
   \hskip 6.35ex $s$, $t$ \hskip 0.4ex
   \raisebox{-1.25ex}[1.0ex][0.0ex]{\hskip 0.0ex {\putlargebrace}}
\else
   {\ifnum\binary > \number\zero
   \hskip 6.35ex $s$, $t$ \hskip 0.4ex
   \raisebox{-3.40ex}[1.0ex][0.0ex]{\hskip 0.0ex {\puthugebrace}}
   \else
   \hskip 6.80ex $s$, $t$ \hskip 0.65ex
   \raisebox{-7.85ex}[1.0ex][0.0ex]{\hskip -2.75ex {\putmammothbrace}}
   \fi}
\fi}
& 49.5 \\
7, 8 & $\rho+\pi\to \rho+\pi$ & $a_{1}(1260)$ & $s$, $t$ & \\
9, 10& $\rho+\pi\to \rho+\pi$ & $\omega(1390)$ & $s$, $t$ & \\
11, 12 & $\rho+K\to \rho+K$ & $K_{1}(1270)$ & $s$, $t$ & 66.0 \\
\\
\multicolumn{5}{c}{$\omega$ meson reactions} \\
1, 2 & $\omega+\pi\to \omega+\pi$ & $b_{1}(1235)$ &
{\ifnum\precount > \number\zero
   \hskip 4.60ex $s$, $t$
   \raisebox{-0.85ex}[1.0ex][0.0ex]{\hskip 1.7ex {\putsmabrace}}
\else
   {\ifnum\binary > \number\zero
   \hskip 4.80ex $s$, $t$
   \raisebox{-2.0ex}[1.0ex][0.0ex]{\hskip 1.7ex {\putmedbrace}}
   \else
   \hskip 5.80ex $s$, $t$
   \raisebox{-5.05ex}[1.0ex][0.0ex]{\hskip 2.4ex {\putbigbrace}}
   \fi}
\fi}
&
{\ifnum\binary > \number\zero
\raisebox{-1.50ex}[1.0ex][0.10ex]{\hskip 0.6ex 74.0}
\else
\raisebox{-2.25ex}[1.0ex][0.10ex]{\hskip 0.6ex 74.0}
\fi}  \\
3, 4 & $\omega+\pi\to \omega+\pi$ & $\rho$ & $s$, $t$ & \\
5, 6 & $\omega+\pi\to \pi+\pi$ & $\rho$ & $s$, $t$ & 21.2 \\
\\
\multicolumn{5}{c}{$\phi$ meson reactions} \\
1 & $\phi+\rho\to K_{1}(1270)+K$ & $K$ & $t$ & 2.29 \\
2 & $\phi+\pi\to K^{*}(892)+K$ & $K$ & $t$ & 6.05 \\
3 & $\phi+K^{*}(892)\to \pi+K$ & $K$ & $t$ & 0.78 \\
4 & $\phi+K_{1}(1270)\to \rho+K$ & $K$ & $t$ & 0.12 \\
5 & $\phi+\rho\to \pi+\rho$ & $\pi$ & $t$ & 0.16 \\
6 & $\phi+\rho\to a_{1}(1260)+\rho$ & $\pi$ & $t$ & 0.07 \\
7,8 & $\phi+\pi \to\pi+\omega$ & $b_{1}(1235)$
& $s$, $t$ & 1.57 \\
9,10 & $\phi+K\to \phi+K$ & $K$ & $s$, $t$ & 3.89 \\
11 & $\phi+\rho\to K+K$ & $K$ & $t$ & 0.34 \\
12 & $\phi+\phi\to K+K$ & $K$ & $t$ & 0.03 \\
13 & $\phi+K^{*}(892)\to K+\rho$ & $\pi$ & $t$ & 0.09 \\
14 & $\phi+K\to K^{*}(892)+\rho$ & $\pi$ & $t$ & 0.85 \\
15 & $\phi+K\to \rho + K$  & $K$ & $t$ & 3.28 \\
16 & $\phi+\pi\to \pi+\pi$  & $\rho(1450)$
& $s$ & 0.58 \\
17 & $\phi+K\to \pi+K^{*}$  & $K$ & $s$ & 7.24 \\
\end{tabular}
\label{table:two}
\end{table}

\begin{references}
\bibitem{remark}Only after computing momentum relaxation and diffusion times
can one be relatively certain about this.
\end{references}
\end{document}